\documentstyle[prb,aps,floats]{revtex}
\input epsf.sty
\newcommand{\rf}[1]{(\ref{#1})}

\newcommand{\jpcm}{J.Phys.: Condens.\ Matter\ }
\begin{document}
%\draft
\twocolumn[\hsize\textwidth\columnwidth\hsize\csname@twocolumnfalse\endcsname

\title{Multiperiodic magnetic structures in Hubbard superlattices}

\author{Andr\'e  L.\ Malvezzi$^{\, (1)}\!$, Thereza Paiva$^{\, (2)}\!$ and
Raimundo R. dos Santos$^{\, (2)}\!$}

\address{$^{(1)}$Departamento\ de F\'\i sica, 
                 Faculdade de Ci\^encias,
                 Universidade Estadual Paulista, 
                 Cx.P.\ 473,
                 17015-970 Bauru SP,
                 Brazil\\
         $^{(2)}$Instituto de F\'\i sica,
                 Universidade Federal do Rio de Janeiro,
                 Cx.P.\ 68528,
                 21945-970 Rio de Janeiro RJ, Brazil}
\date{\today}
\maketitle

\begin{abstract}
We consider fermions in one-dimensional superlattices (SL's), modeled by
site-dependent Hubbard-$U$ couplings arranged in a repeated pattern of
repulsive (i.e., $U>0$) and free ($U=0$) sites.  Density Matrix
Renormalization Group (DMRG) diagonalization of finite systems is used to
calculate the local moment and the magnetic structure factor in the ground
state.  We have found four regimes for magnetic behavior: uniform local
moments forming a spin-density wave (SDW), `floppy' local moments with
short-ranged correlations, local moments on repulsive sites forming
long-period SDW's superimposed with short-ranged correlations, and local
moments on repulsive sites solely with long-period SDW's; the boundaries
between these regimes depend on the range of electronic densities $\rho$
and on the SL aspect ratio. Above a critical electronic density,
$\rho_{\uparrow \downarrow}$, the SDW period oscillates both with $\rho$
and with the spacer thickness. The former oscillation allows one to
reproduce all SDW wave vectors within a small range of electronic
densities, unlike the homogeneous system. The latter oscillation is
related to the exchange oscillation observed in magnetic multilayers. 
A crossover between regimes of `thin' to `thick' layers has also been
observed.  
\end{abstract}
%\bigskip
\vskip2pc]

\section{Introduction}
\label{Intro}

Magnetic multilayers have been the subject of intense study over the last
decade. The technologically important giant magnetoresistance (GMR) is one
of the most interesting aspects of these compounds. Another aspect which
has brought attention to multilayers is the oscillation of the exchange
coupling between magnetic layers as the spacer layer thickness is varied.
While oscillations with single periods have been well understood for some
time, {\em multiperiodicity} has been theoretically predicted,\cite{qw,rkky} 
and indeed observed, in trilayer materials. 
Fe/Cr/Fe samples grown by sputtering or molecular-beam epitaxy (MBE)
display two periods of oscillation of the exchange coupling: a
so-called long-period, of about 10 to 12 monolayers thick, is
superimposed to a short-period component of about two monolayers
thick.\cite{Unguris} This superposition of short- and long-period
components has also been observed in other MBE-fabricated trilayer
materials such as Fe/Mn/Fe,\cite{FeMn} Fe/Au/Fe, \cite{FeAu}
Fe/Mo/Fe,\cite{FeMo} and Co/Cu/Co.\cite{CoCu}
Short-period oscillations, however, disappear if interface quality
is not carefully maintained.\cite{Unguris,Stoeffler,Wolf} Recent
experiments\cite{Schmidt} in Fe/Cr/Fe show that areas of constant Cr
thickness, with diameter larger than 3-4 nm on the interface, are
necessary for the development of short-period oscillations. It is
therefore believed that multiperiodicity has not yet been observed in
multilayers due to interface roughness.

From the theoretical point of view, both the quantum well theory\cite{qw}
and the so-called RKKY theory \cite{rkky} can account for many features
related to the oscillations of the exchange coupling. For instance, a
direct relation between the periods of oscillation and Fermi surface extrema
of bulk spacers has been established.\cite{qw,rkky,mauro} However, since
the notion of a Fermi surface is not widely applicable to strongly
correlated systems, a deeper understanding of multiperiodicity is clearly
in order, and microscopic models should provide useful insights.

With this in mind, here we investigate the magnetic properties of a
one-dimensional superlattice (SL) model\cite{tclp1,longo,mit} in which
electronic correlations are incorporated and treated nonperturbatively. 
The model consists of a periodic arrangement of $L_U$ sites (``layers'') 
in which the on-site coupling is repulsive, followed by $L_0$ free (i.e.,
$U=0$) sites. The role played by relative layer thicknesses on the
magnetic and conducting properties of these systems is then probed by
varying $L_0$ and $L_U$.

The SL structure gives rise to several remarkable features,\cite{tclp1} 
in marked contrast with the otherwise homogeneous system: Local moment
maxima can be transferred from repulsive to free sites, and the range of
parameters in which this occurs has been expressed in terms of a `phase
diagram'.\cite{longo} In addition, spin--density-wave (SDW) quasi-order
can be wiped out as a result of frustration, and the SL structure also
induces a shift in the density $\rho_I$ at which a Mott-Hubbard insulating
phase sets in.\cite{mit} Further, by examining the Luttinger liquid version
of the model,\cite{llsl1} one finds that these superlattices provide the 
means to realize {\em gapless insulating phases.}\cite{llsl2}

Previous studies of the discrete version of the
model\cite{tclp1,longo,mit} resorted to Lanczos diagonalization, which
sets limits on the system sizes used; for instance, a 24-site lattice size
could only be considered for the low- and high-density regimes ($\rho=1/6$
and $\rho=11/6$). Nonetheless, one was still able to probe the period of
exchange oscillations for these special densities through the analysis of
the magnetic structure factor: the peak position displayed oscillatory
behavior with the spacer thickness.\cite{longo} Here we use the Density
Matrix Renormalization Group (DMRG) technique\cite{DMRG} to study
superlattices longer than the ones available through the Lanczos method. 
With the aid of the magnetic structure factor, we have been able to probe
the periodicity of the superlattice over a wider range of layer
thicknesses and densities. As we will see, this has led to significant
improvements on the phase diagram previously reported,\cite{longo} with
the addition of information relative to the regions in which one- and
two--period-oscillations are found; as it turned out, these regions are
closely related to the behavior of the local moment. 
We have also been able to observe a crossover between the regimes of
thin and thick layers; in the latter regime, the `aspect ratio' 
$\ell\equiv L_U/L_0$ is the only relevant geometric parameter, whereas
the magnetic behavior in the former regime depends on $L_U$ and $L_0$
separately.

The layout of the paper is as follows: In Section \ref{model} we introduce
the superlattice model and comment on the calculational procedure. Section
\ref{localmom} focus on the local moment and how it changes with density
and layer thickness.  The magnetic structure factor and the periodicity of
the superlattices are discussed in Section \ref{sq}, and Section
\ref{conc} summarizes our findings.

\section{Model and Calculational Procedure}
\label{model}

We define the Hamiltonian as
\begin{equation}
\label{Ham}
{\cal H}=-t\sum_{i,\ \sigma}
\left(c_{i\sigma}^{^{\dagger}} c_{i+1\sigma}+\text{H.c.}\right)
+ \sum_i U_i\ n_{i\uparrow}n_{i\downarrow}
\label{H}
\end{equation}
where, in standard notation, $i$ runs over the sites of a one-dimensional
lattice, $c_{i\sigma}^{^\dagger}$ ($c_{i\sigma}$) creates (annihilates) a
fermion at site $i$ in the spin state $\sigma=\ \uparrow$ or $\downarrow$,
and $n_i=n_{i\uparrow}+n_{i\downarrow}$, with
$n_{i\sigma}=c_{i\sigma}^{^\dagger}c_{i\sigma}$; the on-site Coulomb
repulsion is taken to be site-dependent: $U_i=U>0$, for sites within the
repulsive layers, and $U_i=0$ otherwise.

We consider the Hamiltonian \rf{H} on lattices with $N_s$ sites and $N_e$
electrons, and open boundary conditions are used.  The appropriate Finite
Size Scaling (FSS) parameter, however, is the number of periodic cells,
$N_c=N_s/N_b$, for a basis with $N_b=L_U + L_0$ sites. The ground state
wave function and energy are obtained by numerical diagonalization using
the finite-system density-matrix renormalization group (DMRG)
method.\cite{DMRG} We used lattice sizes up to 150 sites, and
truncation errors in the DMRG procedure were kept around $10^{-5}$ or
smaller.  We have performed a systematic study of the magnetic properties
for different values of the Coulomb repulsion $U$, different occupation
$\rho = N_e/N_s$ and different configurations $\{U_i\}$. Not all
configurations $\{U_i\}$ fit into all sizes and occupations but, since
DMRG allows us to study a wide range of lattice sizes, we were able to
establish overall trends.

\section{Local Moment Profile}
\label{localmom}

\begin{figure}
\epsfxsize=8.5cm
\begin{center}
%\leavevmode
\epsffile{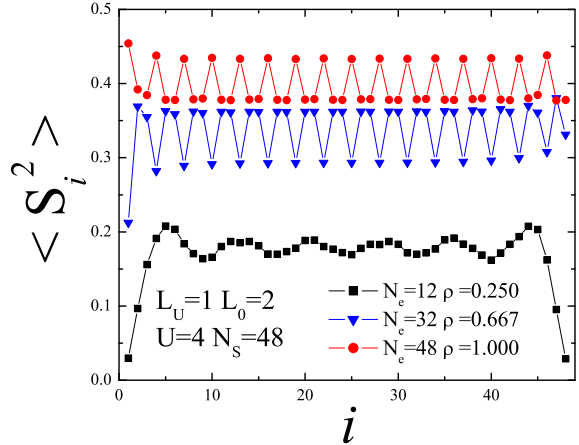}
\caption{Local moment as a function of the site ($i$) for the SL with
$L_U=1$, $L_0=2$, $U=4$, $N_s=48$, for $\rho=0.25$ (squares),
$\rho=0.667$ (triangles) and half-filling (circles). The local
moment profile changes qualitatively as the density increases.} 
\label{si2xi}
\end{center}
\end{figure}

The local moment at site $i$ is defined as $\langle S_i^2\rangle={3\over
4} \langle (n_{i\uparrow}-n_{i\downarrow})^2 \rangle$, and is a measure of
both the magnetism and the degree of itinerancy of the system. Figure
\ref{si2xi} shows the local moment profile for the SL with $L_U=1$,
$L_0=2$, $U=4$, $N_s=48$, and for three different densities; effects of
system size on the local moment are negligible. For small densities, such
as for $\rho=0.25$, one identifies small-amplitude oscillations in the
local moment profile; their period ($2\pi/2k_F$, with $2k_F=\pi\rho$) is
determined not by the underlying SL structure, but by the Friedel
oscillations in the charge density of the otherwise homogeneous
system.\cite{Noack}

As the density is increased, the SL structure dominates over the Friedel
oscillation as evidenced by the data: for $\rho=0.667$ the maxima lie on
the free sites and the modulation of the profile perfectly matches that of
the SL. For large enough densities the maxima migrate to the repulsive
sites, as shown by the data for half filling. One should also note that
even at the maxima $\langle S_i^2 \rangle$ is considerably reduced from
its value at the completely localized limit ($U = \infty$), namely
$\langle S_i^2\rangle=3/4$; the itinerant behavior in these cases is
therefore evident.

The above example illustrates the existence of three regions,
characterized by different local moment profiles: homogeneous (or
Friedel-like), free-site peaked, and repulsive-site peaked.  In order to
locate the boundaries between these regimes it is useful to determine how
the local moment at repulsive and free sites separately change with
the density. In addition, we define a bias of the local moment maxima as
\begin{equation}
\delta \equiv {\langle S_U^2 \rangle - \langle S_0^2 \rangle \over
\langle S_U^2 \rangle +\langle S_0^2 \rangle},
\label{bias}
\end{equation}
and also study its dependence with the density. 

\begin{figure}
\epsfxsize=8.5cm
\begin{center}
%\leavevmode
\epsffile{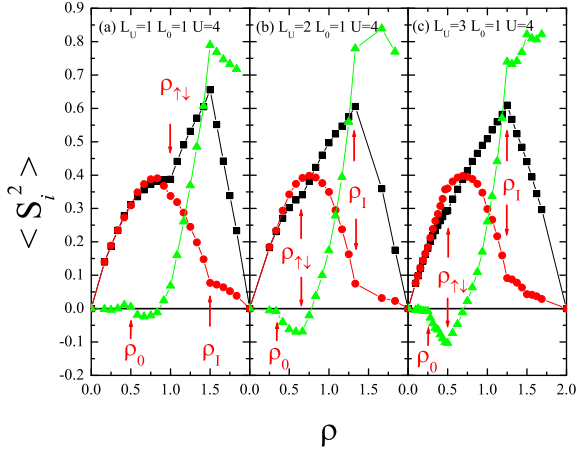}
\caption{Bias (triangles) and local moment at repulsive (squares) and
free (circles) sites, as functions of density, for $U=4$
and for superlattices with (a) $L_U=L_0=1$, $N_s=24$; (b) $L_U=2$,
$L_0=1$, $N_s=48$; and (c) $L_U=3$, $L_0=1$, $N_s=64$.}
\label{l01bias}
\end{center}
\end{figure}

Figure \ref{l01bias} shows the local moment [both at repulsive ($\langle
S_U^2\rangle$) and free sites ($\langle S_0^2\rangle$)] and the bias as
functions of the density, for $U=4$. In the case of Fig.\ \ref{l01bias},
$L_U\geq L_0$, with all SL configurations having $L_0=1$, and $L_U=1$
($N_s=24$), 2 ($N_s=48$) and 3 ($N_s=64$). In order to reduce the effects
of open boundary conditions we have averaged over the 6 innermost cells. 
As the density is increased from the completely empty system, we see that
for densities smaller than $\rho_0$, given by
\begin{equation} 
\rho_0 ={1 \over {L_0+L_U}},
\label{rho0}
\end{equation}
the local moment increases, and is the same on both sublattices (hence
$\delta=0$). For the SL's with $L_0=1$ and $L_U=1$, 2 and 3 one has 
$\rho_0=0.5$, 0.33 and 0.25, respectively, which are indicated by 
arrows in Fig.\ \ref{l01bias}. 
This density corresponds to having one electron on each
cell, so that in the case $L_U\geq L_0$ electrons have equal probability
of being either on a free or on a repulsive site for $\rho < \rho_0$.

 % Fig 2

From Fig.\ \ref{l01bias} one sees that there is a range of densities above
$\rho_0$, in which the local moment grows slower on the repulsive sites
than on the free ones, since added electrons will preferentially occupy
the free sites; hence a negative bias develops within this range. By the
same token, $\langle S_0^2 \rangle$ will reach its maximum value at
densities smaller than those at which $\langle S_U^2 \rangle$ displays its
maximum; for completeness, recall that the maximum value of
the local moment on a homogeneous free lattice is 0.375, occurring at half
filling.

In strong coupling, the free sites saturate at the density
\begin{equation}
\rho_{\uparrow \downarrow} ={{2L_0} \over {L_0+L_U}},
\label{rhoud}
\end{equation}
which corresponds to having two electrons on each free site, while the
repulsive site is empty. Nonetheless, even for moderate couplings, this
density is special. Indeed, from Figs.\ \ref{l01bias}(b) and
\ref{l01bias}(c) one can see that for $L_U > L_0$ the bias reaches its
minimum value exactly at $\rho_{\uparrow \downarrow}$.  Also, the local
moment at repulsive sites shows a bump at $\rho_{\uparrow \downarrow}$, 
indicating the beginning of a steady occupation of repulsive sites.

\begin{figure}
\epsfxsize=8.5cm
\begin{center}
%\leavevmode
\epsffile{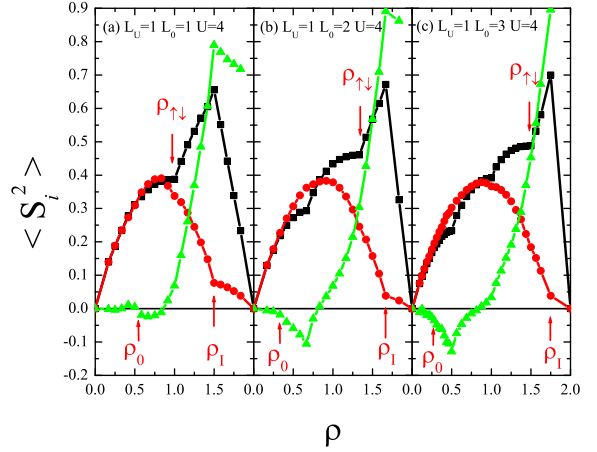}
\caption{Bias (triangles) and local moment at repulsive (squares) and
free (circles) sites, as functions of density, for $U=4$
and for superlattices with (a) $L_U=L_0=1$, $N_s=24$; (b) $L_U=1$,
$L_0=2$, $N_s=48$; and (c) $L_U=1$, $L_0=3$, $N_s=64$.}
\label{lu1bias}
\end{center}
\end{figure} % Fig 3

Increasing the density even further, one sees that $\langle S_U^2\rangle$
reaches its maximum at $\rho_I$, defined as
\begin{equation}
\rho_I ={{2L_0+L_U} \over {L_0+L_U}},
\label{rhoi}
\end{equation}
which, in strong coupling, corresponds to having two electrons on each
free site and one on each repulsive site; the maximum of $\langle
S_U^2\rangle$, at exactly this density, is indicative of the SL being in a
Mott-Hubbard insulating state.\cite{mit} In the region between
$\rho_{\uparrow \downarrow}$ and $\rho_I$, the repulsive layer is
preferentially filled as the overall density is increased, causing a steep
rise (drop) in the local moment at the repulsive (free) sites. As Fig.\
\ref{l01bias} shows, the consequence is a steady increase of the bias in
this interval.

For densities larger than $\rho_I$ the free sites are almost completely
filled, which is apparent by a considerable decrease in the magnitude of
the derivative of $\langle S_0^2\rangle$ with respect to the density.
Fermions will then start to double occupy the repulsive sites, thus
causing a reduction in $\langle S_U^2\rangle$.

The equivalent of Fig.\ \ref{l01bias} for the case $L_U\leq L_0$ is shown
in Fig.\ \ref{lu1bias}. While the overall behavior is the same,
a few differences are worth mentioning. The first one is the behavior of
the bias in the range $\rho < \rho_0$: While the bias vanishes for
$L_U\geq L_0$, for $L_U < L_0$ it is negative, though of small magnitude. 
This is due to the fact that in this range of densities, and within each
cell, the electrons have more free sites at their disposal to resonate
than repulsive ones; this excess of free sites within each cell also
explains why the bias still decreases for densities above $\rho_0$. 

Second, for $L_U < L_0$, $\langle S_U^2 \rangle$ is boosted whenever
$\rho=2m\rho_0$, with $m=1,2,\ldots,L_0+L_U-1$; this can be attributed to
the fact that the double occupancy of the repulsive sites is least likely
whenever there are an even number of electrons per cell. Note also that
the first bump, at $2\rho_0$, coincides with the minimum of the bias. 

And, third, while for $L_U > L_0$ the bias changes sign for
$\rho_{\uparrow \downarrow}< \rho < \rho_I$, when $L_U \leq L_0$ this
occurs for $\rho_0 < \rho < \rho_{\uparrow \downarrow}$; the actual
location of the density at which $\delta=0$ depends on the SL
configuration, as well as on the Coulomb repulsion $U$.

\begin{figure}[t]
\epsfxsize=8.5cm
\begin{center}
%\leavevmode
\epsffile{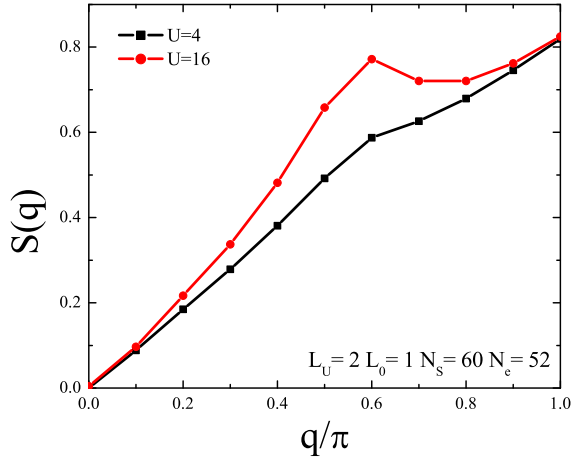}
\caption{Magnetic structure factor [Eq.\ (\ref{sofq})] for a SL with
$L_U=2$,
$L_0=1$, $N_s=60$, 
$N_e=52$ (hence $\rho=0.87$), and for $U=4$ and $16$.} 
\label{squ}
\end{center}
\end{figure} % Fig 4

\section{Magnetic Structure Factor and effective densities}
\label{sq}

Let us now turn to the magnetic structure factor, which is defined as
\begin{equation}
{\cal S}(q)={1\over N_c} \sum_{i,j} {\rm e}^{iq(r_i-r_j)}
\langle {\bf S_i}\cdot {\bf S_j}\rangle.
\label{sofq}
\end{equation}
As $q$ is related to
the repeating units, ${\cal S}(q)$ probes the relative arrangement between
different cells. It is important to have in mind that the homogeneous
system displays a single peak in the magnetic structure factor at
$q_{max}= 2 k_F = \pi \rho$, for $ \rho \le 1$, or $q_{max}=2 k_F= \pi (2-
\rho)$, for $\rho \ge 1$;\cite{Frahm90} the lattice spacing is taken to be
unity throughout this paper. 

\begin{figure}
\epsfxsize=8.5cm
\begin{center}
%\leavevmode
\epsffile{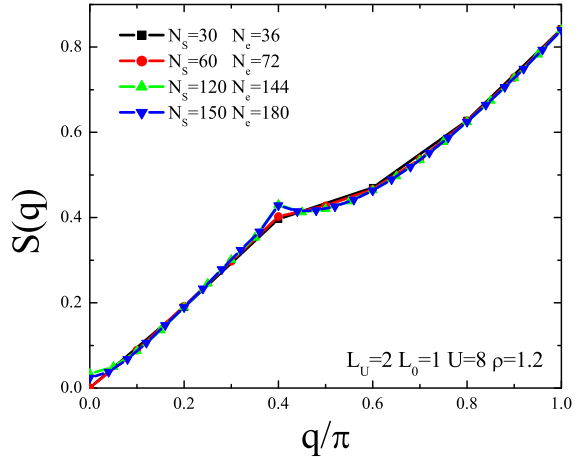}
\caption{Magnetic structure factor [Eq.\ (\ref{sofq})] for a SL with
$L_U=2$, $L_0=1$, $U=8$, 
$\rho=1.2$, and for different system sizes: $N_s=30$ (squares), $N_s=60$
(circles), $N_s=120$ (up triangles), and $N_s=150$ (down triangles).}
\label{fse}
\end{center}
\end{figure} % Fig 5

Figure \ref{squ} shows ${\cal S}(q)$ for a SL with $L_U=2$, $L_0=1$,
$\rho=0.87$, and for two values of $U$, namely $U=4$ and $U=16$. Two peaks
in the magnetic structure factor are clearly seen in this case: one at
$q=\pi$, and another at $q=3 \pi/ 5$. While the former is not affected by
an increase in $U$, the latter grows with $U$, though without changing its
position. Actually, for sufficiently large $U$ the peak at $q\neq \pi$
even becomes more pronounced than the one at $q=\pi$. 
Further data
show that this happens for a range of values of $L_0$, $L_U$, and $\rho$,
as discussed below.

The presence of two peaks (at, say, $q_{max}$ and $q_{max}'$, with
$q_{max}'<q_{max}$) in the structure factor is associated with a tendency
of the system to order (strictly speaking, to {\em quasi-}order, in one
dimension) in a magnetic arrangement dominated by the corresponding
periods, $\lambda=2\pi/q_{max}$ and $\lambda'=2\pi/q_{max}'$.  As we will
see below, the long period oscillates with the spacer thickness, a
behavior reminiscent of the exchange oscillation observed in magnetic
trilayers. 

These two peaks also differ in the way they depend on the system size. 
Figure \ref{fse} shows ${\cal S}(q)$ for the SL with $L_U=2$, $L_0=1$,
$U=8$, $\rho=1.2$ ($\rho_{\uparrow \downarrow} < 1.2 <\rho_I$), and for
four different lattice sizes, ranging from $N_s=30$ to $N_s=150$. From
Fig.\ \ref{fse} we see that the inflection already present for $N_s=30$ at
$q= 2\pi/5$ sharpens as $N_s$ increases, and that there is no change in
the position of the peak. We have checked that a similar slow, but steady,
growth of the peak height with $N_s$ occurs for the homogeneous Hubbard
model away from half filling.  These features have been observed for other
SL configurations and densities, which therefore indicate that whenever a
peak is found at $q \ne \pi $, it is robust. On the other hand, the peak
at $q=\pi$ shows a much weaker size-dependence, so that it should be
associated with strong, although short-ranged, correlations; this point is
illustrated below.

We can then turn to a systematic study of the number of peaks and their 
positions, by analysing
the evolution of the structure factor as the density of electrons is
increased.  As discussed in Sec.\ \ref{localmom}, for $\rho < \rho_0$ the
local moments are small, and either their maxima are on the free layers,
or they are evenly distributed throughout the lattice, depending on
whether $L_U < L_0$ or $L_U \ge L_0$, respectively.  A small or zero bias
signals that the SL structure is not very relevant in this situation.
Indeed, the spatial decay of the spin-spin correlation function (not shown
here) in the case of a superlattice with a small bias can hardly be
distinguished from that of the corresponding homogeneous system; as a
result, the magnetic structure factor displays a single peak.
In addition, this single peak displays a size- and a $U$-dependence 
similar to those for the homogeneous system.

\begin{figure}[t]
\epsfxsize=8.5cm
\begin{center}
%\leavevmode
\epsffile{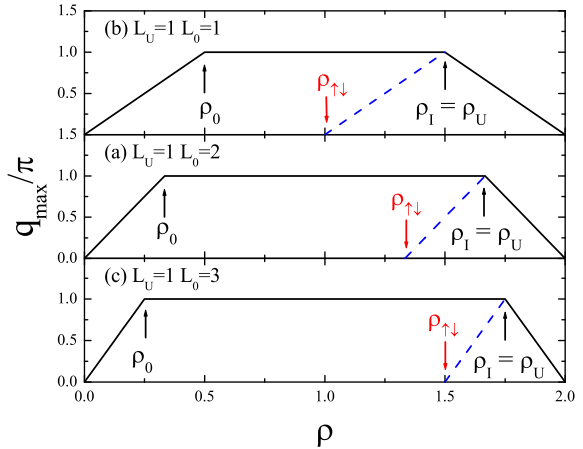}
\caption{Maxima position, $q_{max}$, of the magnetic structure factor as
a function of density for $U=4$, and
(a) $L_U=1$ $L_0=1$ $N_s=24$,
(b) $L_U=1$ $L_0=2$ $N_s=48$, and (c) $L_U=1$ $L_0=3$ $N_s=64$. The dashed
lines indicate the presence of another peak in ${\cal S}(q)$ (see text).}
\label{qmaxlu1}
\end{center}
\end{figure} % Fig 6

\begin{figure}
\epsfxsize=8.5cm
\begin{center}
%\leavevmode
\epsffile{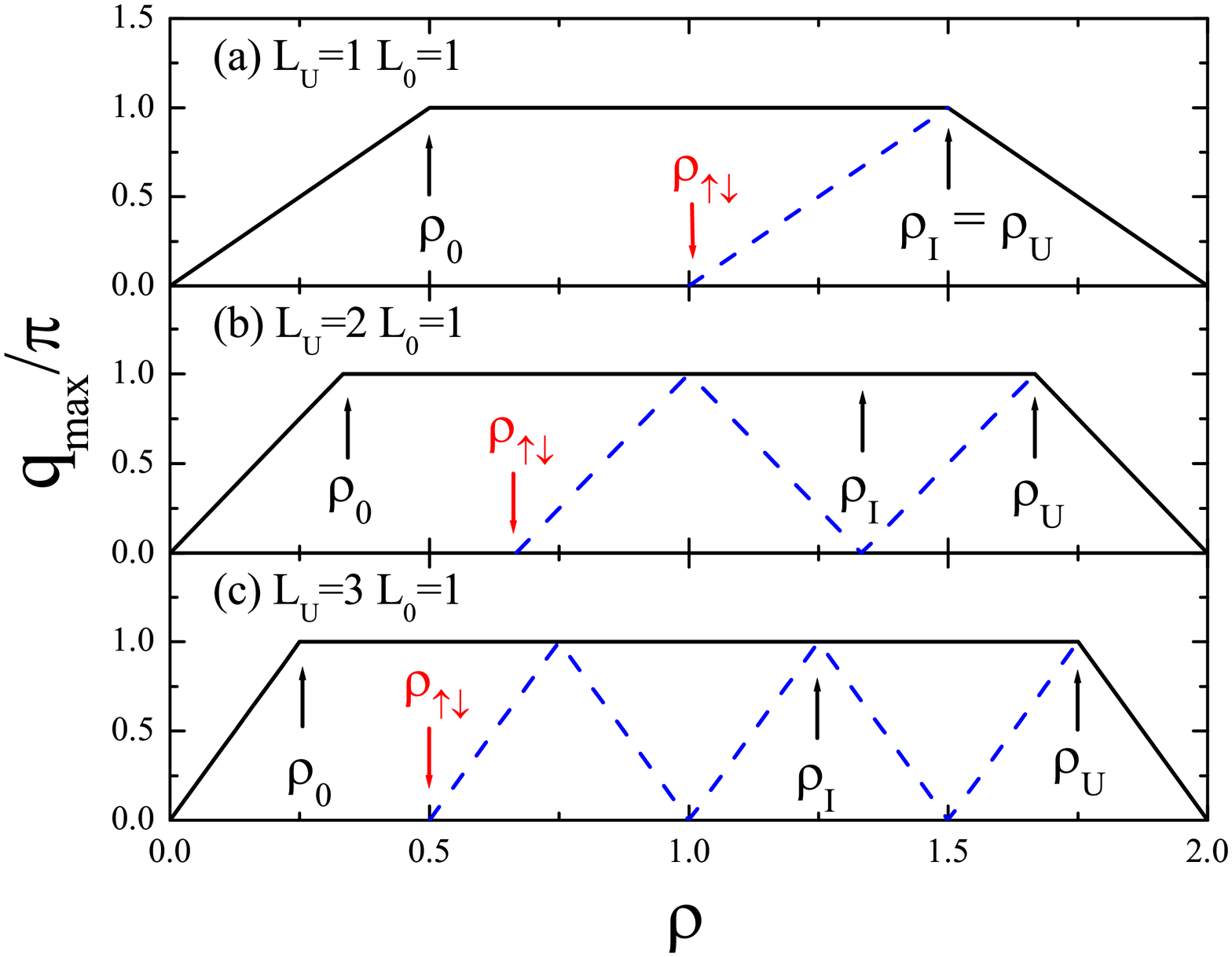}
\caption{Maxima position, $q_{max}$, of the magnetic structure factor as a
function of density for $U=4$, and
$L_U=1$ $L_0=1$ $N_s=24$ (a),
$L_U=2$ $L_0=1$ $N_s=48$ (b), and 
$L_U=3$ $L_0=1$ $N_s=64$ (c). The dashed
lines indicate the presence of another peak in ${\cal S}(q)$ (see text).}
\label{qmaxl01}
\end{center}
\end{figure}

In order to relate the peak position with some density, one can 
think of a free (homogeneous) lattice in which the sites are grouped 
in cells mimicking the SL structure under consideration; 
it then follows that a meaningful quantity is the {\em cell 
density} of electrons, 
\begin{equation}
\rho_{\rm cell}={N_e\over N_c}=\rho(L_U+L_0),
\label{rhocell}
\end{equation} 
where $\rho$ is the overall density.
For the interacting SL, we have found that the peak position 
is given by the same 
expression as for the homogeneous case, but with $\rho_{\rm cell}$ 
replacing $\rho$; that is, 
\begin{equation}
q_{max}= \pi \rho_{\rm cell},\ \ \text{for}\ \rho\leq\rho_0.
\label{qm1}
\end{equation}
Thus, the peak position grows linearly with $\rho$ up to $\rho=\rho_0$ 
(at which density $\rho_{\rm cell}=1$), when it reaches $q_{max}= \pi$; 
see Figs.\ \ref{qmaxlu1} and \ref{qmaxl01}.

 % Fig 7

\begin{figure}
\epsfxsize=8.5cm
\begin{center}
%\leavevmode
\epsffile{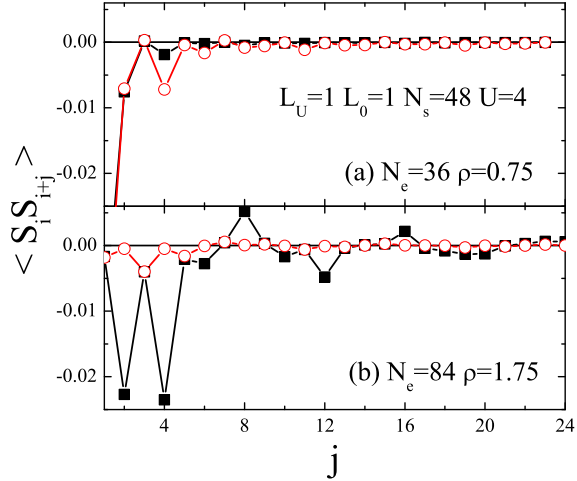}
\caption{Spatial decay of correlations, for a SL with $L_U=L_0=1$ 
and $U=4$
for $N_s=48$ sites: in (a) $\rho =0.75$, and in (b) $\rho=1.75$.
Circles and squares correspond to the origin being taken on free 
and repulsive sites, respectively.}
\label{sofr}
\end{center}
\end{figure}

The single peak regime persists for $\rho_0 < \rho < \rho_{\uparrow
\downarrow}$, and, as shown in Figs.\ \ref{qmaxlu1} and \ref{qmaxl01}, 
now the
peak is always at $q=\pi$. The single peaks in this region show a very
weak dependence on the system size, which is reflected in the spatial
decay of the correlation functions, $\langle S_0^zS_j^z\rangle$.  As
illustrated in Fig.\ \ref{sofr}(a) for $L_U=L_0=1$, $U=4$, and
$\rho=0.75$,
correlations with origin in either of the sublattices barely survive at
large distances; this should be contrasted with the case displayed in
Fig.\ \ref{sofr}(b), for $\rho=1.75$ (see below), in which correlations 
in one of the superlattices are `long' ranged.

At  $\rho_{\uparrow \downarrow}$, and in strong coupling, the free 
layers are completely filled while the repulsive layers are empty. 
But as the density is increased beyond $\rho_{\uparrow \downarrow}$, 
a second peak emerges, 
%whose position, $q_{\rm max}'$, grows from $q=0$, 
as indicated by the dotted lines in Figs.\ \ref{qmaxlu1} and \ref{qmaxl01}. 
This second peak results from the robust moments located
on the repulsive sites. Indeed, if one defines an effective 
electronic density on the {\em repulsive} layers as
\begin{equation}
\rho_{\rm eff}=\rho(L_0+L_U)-2L_0,
\label{rhoeff}
\end{equation}
where $\rho$ is the overall density, the long-periods are located at
\begin{equation}
q_{max}'=\pi\rho_{\rm eff}.
\label{qmax}
\end{equation}
With this definition, it also becomes clear that for 
$\rho=\rho_{\uparrow \downarrow}$ there is no net moment at the repulsive
layers, since $\rho_{\rm eff}=0$.

 % Fig 8

This two peak structure is present until one reaches $\rho_U$, defined by 
\begin{equation}
\rho_U \equiv 2-\rho_0,
\label{rhoU}
\end{equation}
which corresponds to a single hole per cell.
One should also have in mind that the overall magnetic arrangement is 
determined by the long-period (characterized by $q_{max}'$), since, 
as discussed above, this is the one increasing with system size.

An interesting difference between the cases depicted in Figs.\
\ref{qmaxlu1} and \ref{qmaxl01} is the fact that in the
former $\rho_U=\rho_I$, while in
the latter $\rho_U>\rho_I$, and $q_{\rm max}'$ is able to go
through at least
one complete oscillation before $\rho$ reaches $\rho_U$.
In this case, the situation $q_{max}'=0$ does not indicate any tendency 
towards a ferromagnetic arrangement, but is to be associated with 
{\em frustration} of the corresponding long-period SDW.\cite{longo} 
Indeed, when $L_U=2$ and $L_0=1$ the
Mott-Hubbard insulator at $\rho_I$ is frustrated, since two spins on each
repulsive layer form local singlets. Singlets on different repulsive
layers, in turn, do not couple with each other, though short ranged
correlations are still present; see Fig.\
\ref{qmaxl01}(b).  The frustration at half filling for $L_U=3$ and $L_0=1$
can be understood by a similar strong coupling analysis: of the four
electrons on each cell, two occupy the free site and the remaining two
resonate between three sites, but always forming a singlet. Figure
\ref{qmaxl01}(c) shows that further addition of electrons renders these
singlets unfavorable, and the system again displays a SDW. At $\rho=3/2$,
one reenters a frustrated state, again as a result of having an even number of
electrons on the repulsive layer. Therefore, we can relate the
reentering frustrated configurations to the formation of singlets on
the repulsive layer, which occurs whenever there is an even number of
electrons per cell; that is, whenever the density goes through an even
multiple of $2\rho_0$.

And, finally, above $\rho_U$ all SL's return to a single peak regime: 
${\cal S}(q)$ has a maximum at $\pi(2-\rho_{\rm cell})$ [$=\pi(2-\rho_{\rm
eff})$, since $\rho_{\rm eff}=\rho_{\rm cell}-2L_0$]. The correlations 
in this regime are quasi--long-ranged, since, as shown in Fig.\ 
\ref{sofr}(b), the correlation function with origin at a repulsive site is
slowly decaying.

\begin{figure}[t]
\epsfxsize=8cm
\begin{center}
%\leavevmode
\epsffile{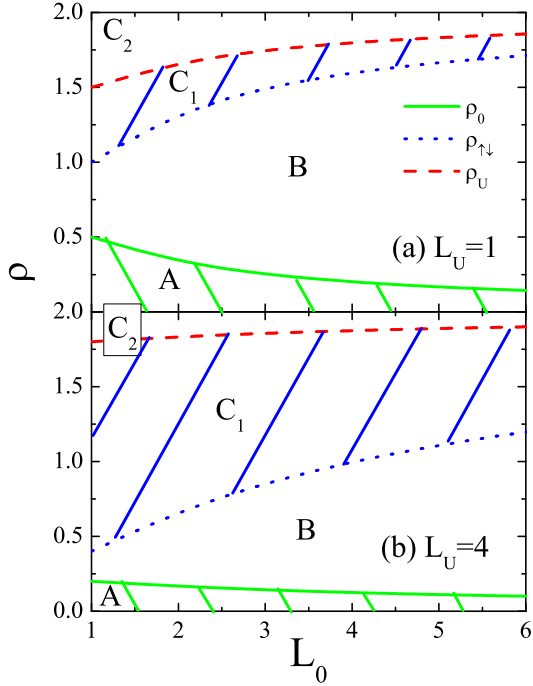}
\caption{Regions in the parameter space $(\rho,L_0)$ for (a) $L_U=1$ and
(b) $L_U=4$:  A -- weak moments formed preferentially on free layers and
one single SDW period; B -- local moment maxima depend on $U$ and on
$L_0/L_U$ and spin correlations are predominantly antiferromagnetic, but
short ranged; C1 -- local moment maxima on the repulsive layers and the
SDW's are dominated by two periods; $C_2$ -- local moment maxima on the
repulsive layers and the SDW's are dominated by a single period. See text
for details. Full line corresponds to $\rho_0$, dotted line to
$\rho_{\uparrow\downarrow}$ and dashed line to $\rho_U$.}
\label{diag2d}
\end{center}
\end{figure}

The above analyses of the magnetic structure factor and of the local
moment profile can be extended to several other SL configurations, and the
outcome can be best summarized by a diagram in the parameter space
$(\rho,L_0,L_U)$, showing the presence of four different regions (or
phases). Cross sections of the full three-dimensional phase diagram are
presented in Fig.\ \ref{diag2d}(a) for $L_U = 1$ and Fig.\ \ref{diag2d}(b) 
for $L_U=4$.  In the low density region (A), located between $\rho=0$ and
$\rho=\rho_0$, the system behaves roughly as if it were homogeneous. The
local moments are small and their maxima are located preferentially on the
free layers. The SDW is dominated by a single density-dependent
wavevector, $q_{max}=\pi\rho_{cell}$. 

 % Fig 9 

At somewhat larger densities, $\rho_0 < \rho <\rho_{\uparrow \downarrow}$,
lies a region (B), in which the positions of the maxima in the local
moment profile depend on the repulsion $U$, on $L_0$, and $L_U$. 
Presumably as a result of this `floppy' character of the local moments,
spin correlations in this region are strongly antiferromagnetic, but
short-ranged.

As the density is further increased, one enters the double-period region
($C_1$), which lies between $\rho=\rho_{\uparrow \downarrow}$ and
$\rho=\rho_U$. In this region the local moment on the repulsive layer
suffers successive boosts, and one finds SDW's with a `long' period
$\lambda'=2/\rho_{\rm eff}$; the latter are accompanied by strong short
ranged correlations, of period $\lambda=2$. 

And, finally, there is a high density region ($C_2$), with densities
ranging from $\rho=\rho_U$ to $\rho=2$. At $\rho_U$ the local moment bias
is maximum (see Figs.\ \ref{l01bias} and \ref{lu1bias}), so it decreases
as one increases the density. Nonetheless, one still has SDW's, now with a
single period given by $\lambda=2/(2-\rho_{\rm cell})$. By comparing the
two cases depicted in Fig.\ \ref{diag2d}, one sees that a growth of the 
repulsive layer increases the two-peaked region, at the expense of all
others. 

\begin{figure}
\epsfxsize=8.5cm
\begin{center}
%\leavevmode
\epsffile{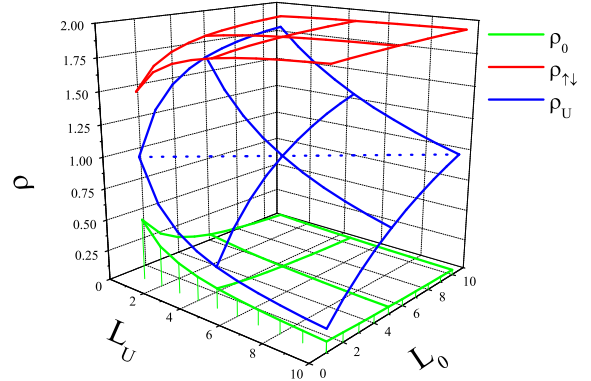}
\caption{Regions in the parameter space $(\rho,L_0,L_U)$:  the lower
surface corresponds to $\rho_0$, the middle one to
$\rho_{\uparrow\downarrow}$, and the upper one to $\rho_U$.
The dotted line is the intersection of the
$\rho_{\uparrow \downarrow}$-surface with the plane $\ell=1$ (see text).}
\label{diag3d}
\end{center}
\end{figure}

The full three-dimensional phase diagram is shown in Fig.\ \ref{diag3d}. 
The densities $\rho_0$, $\rho_{\uparrow \downarrow}$, and $\rho_U$ define
surfaces in the parameter space $(\rho,L_0,L_U)$ which act as boundaries
between the four regions discussed above. The $\rho_0$-surface flattens
considerably for thick layers, and if one imagines an $\ell=1$ line on the
horizontal plane of the figure, we see that the homogeneous-like region is
only important for moderately thin layers. The $\rho_{\uparrow
\downarrow}$-surface is the same whether the layers are short or long,
since it depends on $L_U$ and $L_0$ only through the combination
$L_U/L_0\equiv\ell$; to illustrate this, the intersection of the
$\rho_{\uparrow \downarrow}$-surface with the plane $\ell=1$, shown as a
dotted line in Fig.\ \ref{diag3d}, yields $\rho_{\uparrow \downarrow}=1$
for all $L_U=L_0$.  The topmost surface ($\rho_U$) also displays a similar
crossover between thin and thick regimes: for thick lattices $\rho_U\to
2$. 

 % Fig 10  

We are now in a position to discuss the oscillation in $q_{max}$ with the
spacer (free layer) thickness, for a {\em fixed electron density}; as
mentioned before, these are related to the oscillation of the exchange
coupling between magnetic layers. 
When $\rho \leq \rho_0$, one has $q_{max}\leq\pi$, so that there is no
oscillatory behavior in $q_{max}$. For $\rho_0 < \rho \leq
\rho_{\uparrow\downarrow}$, the peak is always at $q_{max}=\pi$, so that
again no oscillation is found. Above $\rho_{\uparrow\downarrow}$, the
long-period 
maxima in the magnetic structure factor are located at
$q_{max}(L_0)=\pi
\rho_{\rm eff}(L_0)$, where we have emphasized the dependence with $L_0$
through $\rho_{\rm eff}$. We can then calculate the period of oscillation,
$\Delta L_0$, by setting $q_{max}(L_0)=q_{max}(L_0+\Delta L_0),\ {\rm
mod}(2\pi)$. For $\rho_{\uparrow\downarrow} < \rho \leq 1$ we get
\begin{equation}
\Delta L_0 = \left(1-{k_F \over \pi}\right)^{-1},
\label{period1}
\end{equation}
where $2k_F=\pi\rho$ since $\rho\leq 1$; for $1<\rho\leq\rho_U$, we
similarly find
\begin{equation}
\Delta L_0 = {\pi\over k_F},
\label{period2}
\end{equation}
where now $2k_F=\pi(2-\rho)$.  Note that this result is not valid for
$L_U=1$, since, according to Fig.\ \ref{qmaxl01}, once $q_{max}$ vanishes,
it does not grow as $L_0$ increases. For $\rho>\rho_U$, on the
other hand, Eq.\ (\ref{period2}) is applicable.

We have then established that (i) $\rho_{\uparrow\downarrow}$ acts as a
critical density for the appearance of `exchange oscillations', and that
(ii) our previous results\cite{longo} for $\Delta L_0$, obtained in the
high density region, are valid quite generally for $\rho >
\rho_{\uparrow\downarrow}$. Further, Eqs.\ (\ref{period1}) and
(\ref{period2})  reproduce previous findings, within the Hartree-Fock
approximation, for the periods of oscillation of the exchange coupling in
magnetic multilayers.\cite{qw,mauro} Also, the experimentally
observed short period of two
monolayers reported in Ref.\ \onlinecite{Unguris} corresponds, in our
framework, to the $\lambda = 2$ correlations.  Thus, electronic
correlations do not modify the quantum interference effects determining
the periods of oscillation from the extrema of the Fermi surface of the
spacer material.

\section{Conclusions}
\label{conc}

We have investigated one-dimensional Hubbard superlattices consisting of
periodic arrangements of free and repulsive layers, by means of Density
Matrix Renormalization Group. By considering a much wider range of lattice
sizes and densities than in previous studies, we have refined in several
aspects our earlier predictions for the magnetic behavior. There are now
{\em four} distinct regimes, depending on the range of electronic
densities.  For less than one electron per periodic cell, the local moment
profile is approximately uniform, and spin-density waves are dominated by
a single density-dependent wave vector. When the density lies between
those corresponding to one electron per cell and to a fully occupied free
sublattice (with empty repulsive sites), maxima in the local moment
profile develop, which can be either on free sites or on the repulsive
sites, depending on the SL configuration, on the density, and on $U$; 
also, spin correlations become short ranged, but dominated by a tendency
of neighboring cells to align antiparallel. For densities larger than two
electrons per free site one has a two-period magnetic structure. There is
a long period SDW, in which the wave vector oscillates as a function of
the electronic density.  An immediate consequence is that SDW's with all
possible wave vectors are generated within an interval of densities of
$2\rho_0$; this should be compared with the homogeneous system, for which
one needs to vary between an empty lattice and a half filled one in order
to generate all possible wave vectors.  These long-period SDW's are
superimposed with short ranged correlations with $q=\pi$, which disappear
for densities above one hole per periodic cell.

We have also extended to a broader range of densities our earlier
prediction that the wave vectors for the SDW's oscillate as the free layer
length is varied, with a period determined solely by the electronic
density (through the Fermi wave vector). 
In the context of magnetic multilayers, our results for the period of
oscillations exactly reproduce the relation between Fermi
surface extrema with exchange coupling oscillation; in addition, the
two-monolayer period observed experimentally corresponds in our model to
the short period at $q=\pi$.  

And, finally, we have been able to observe a
crossover between the regimes of thin and thick layers; in the latter, the
`aspect ratio' $\ell\equiv L_U/L_0$ is the only relevant geometric
parameter, whereas in the former regime the magnetic behavior depends on
$L_U$ and $L_0$ separately. For instance, when any of the layers are thin
-- less than about 6 sites long --, the SL structure is not felt at low
densities, and it behaves as if it were homogeneous; for thick
layers, this quasi-homogeneous behavior is only noticeable at very low
densities. Similarly, the region of singly peaked correlations at high
densities gets smaller as the layers get thicker. 

As a final comment, one should expect that the applicability of the
one-dimensional model treated here is very close to being extended beyond
the realm of higher dimensional superlattices. Indeed, fabrication of
nanowire superlattices has been recently reported.\cite{Gudiksen02}
Although these superlattices were made up of semiconducting materials, the
prospects of growing metallic and/or magnetic {\em nanosuperlattices} are
promising. In this case, our results indicate that a careful control of
the doping level leads to a wide variety of distinct magnetic behaviors in
the same material. Another possible realization of our model would be to a
(as yet hypothetical) superlattice made up of single-walled metallic
carbon nanotubes, since these have been successfully described in terms of
a Luttinger liquid; see, e.g., Ref.\ \onlinecite{llsl2} for a partial list
of references.

\acknowledgments

The authors are grateful to J.~d'Albuquerque e Castro, E.~Miranda, and
J.~Silva-Valencia for discussions. Financial support from the Brazilian
Agencies CNPq (ALM and RRdS), Funda\c c\~ao de Amparo \`a Pesquisa do
Estado de S\~ao Paulo-FAPESP (ALM), FAPERJ (TP and RRdS), and `Millenium
Institute for Nanosciences/CNPq-MCT' (RRdS) is also gratefully
acknowledged.

\end{document}